# Using Text Embeddings for Deductive Qualitative Research at Scale in Physics Education

Tor Ole B. Odden[1], Halvor Tyseng[2,†], Jonas Timmann Mjaaland[2,†], Markus Fleten Kreutzer[2,†], Anders Malthe-Sørenssen[1,2]

[1]Center for Computing in Science Education, University of Oslo, 0316 Oslo, Norway
[2]Center for Interdisciplinary Education, University of Oslo, 0316 Oslo, Norway

†*These authors contributed equally to this work.*

## ABSTRACT

We propose a technique for performing deductive qualitative data analysis at scale on text-based data. Using a natural language processing technique known as text embeddings, we create vector-based representations of texts in a high-dimensional meaning space within which it is possible to quantify differences as vector distances. To apply the technique, we build off prior work that used topic modeling via Latent Dirichlet Allocation to thematically analyze 18 years of the Physics Education Research Conference proceedings literature. We first extend this analysis through 2023. Next, we create embeddings of all texts and, using representative articles from the 10 topics found by the LDA analysis, define centroids in the meaning space. We calculate the distances between every article and centroid and use the inverted, scaled distances between these centroids and articles to create an alternate topic model. We benchmark this model against the LDA model results and show that this embeddings model recovers most of the trends from that analysis. Finally, to illustrate the versatility of the method we define 8 new topic centroids derived from a review of the physics education research literature by Docktor and Mestre (2014) and re-analyze the literature using these researcher-defined topics. Based on these analyses, we critically discuss the features, uses, and limitations of this method and argue that it holds promise for flexible deductive qualitative analysis of a wide variety of text-based data that avoids many of the drawbacks inherent to prior NLP methods.

## I. INTRODUCTION

Qualitative analysis is not easy. It is time-consuming, and often difficult to replicate. In physics education research (PER) specifically, many qualitative studies aim to infer students' or instructors' understandings, framings, emotions, or social dynamics, and therefore require fine-grained analysis of subtle details of dialogue, gesture, or speech. Categorizations can hinge on the use of specific phrases or words, the lack thereof, or even require researchers to go beyond surface content and context to focus on more gestalt aspects of the data.

Despite these challenges, qualitative analysis is often the best method we have for many of the research problems in a field as diverse, multi-faceted, and complex as PER. Ours is a fairly new field, and as such is still in the process of identifying, exploring, and drawing in various methods and research questions [1]. It also addresses a sufficiently complex topic— human learning—that the methods used often look less like standard physics than biology research, where qualitative methods are much more prevalent [2].

At the same time, there is plenty of room for methodological improvement. As theoretical paradigms and research questions become established, researchers often move from theoretical papers, small-N case studies, and existence proofs (i.e., [3,4]) to large-scale validation studies (i.e., [5,6]). But large qualitative studies often come at the cost of either significantly higher time investment or lower analytical grain size.

Over the past several years, the field of natural language processing (NLP) has made huge gains in the mathematical representation, manipulation, and analysis of text. These tools hold significant promise in supporting qualitative research on text-based data, since computers are now able to reliably perform many of the types of content-based analyses that were previously only possible to do by a human researcher. Thus, these methods have the potential to help researchers scale up their qualitative analyses. In this article we show a proof-of-concept of how one of these current, cutting-edge NLP methods can be used as an aid in performing deductive qualitative analysis at scale in PER.

## II. THEORY

### A. Qualitative research in PER: deductive and inductive approaches

Qualitative research is generally divided into two approaches: deductive and inductive. When using a deductive approach, researchers typically analyze a dataset using a pre-established categorization scheme: often a set of codes, categories, or a theoretical framework that defines the different categories used in the analysis. These codes may be mutually exclusive (i.e., each piece of data can only be given one code) or overlap; they may be applied as a binary, or using different levels of prevalence, confidence, or saturation. Once the data is coded, the codes may be used to organize or summarize the data [7], or form the basis for further quantitative analysis based on the prevalence and co-occurrence of codes in the data.

Inductive approaches, in contrast, allow researchers to make sense of data by deriving or discovering patterns, trends, or themes. Depending on the particular method, researchers may go into the inductive analysis with some ideas of what they are looking for or try to avoid bringing in any a-priori ideas and let the data "speak for itself" (a variant known as grounded theory). They then review the data multiple times, building and refining their themes or categorizations with each pass through the dataset. Once the data is analyzed, the results can be used to describe theoretical or empirical phenomena, propose new frameworks (or variants of existing frameworks), or explain social dynamics or lived experiences.[1]

In PER specifically, both deductive and inductive approaches are common [8]. Many of the foundational studies of student understanding of physics were based on interviews with students, which were by necessity analyzed qualitatively [1,9,10]. Even quantitative instruments to measure student attitudes or understandings, like the Force Concept Inventory [11], typically involve some amount of qualitative data analysis to establish which ideas or attitudes the instrument is to probe [12]. However, the actual process of qualitative data analysis often looks much the same today as it did 20 or even 40 years ago, aside from some digital advances in data storage and qualitative data analysis software.

Over the last years, the fields of artificial intelligence (AI) and natural language processing (NLP) have seen huge advances with the development of large language models and generative AI. Much of this advancement is based on increasingly sophisticated ways of representing text as vectors, which allows the text to be mathematically manipulated using

---

[1] For a thorough description of the qualitative research process in PER, we recommend Otero, Harlow, & Meltzer (2023) [8]

standard statistical and vector-mathematical methods [13–15]. Thus, we see a need to explore how these tools may help researchers qualitatively analyze their data.

### B. NLP analysis in PER—supervised and unsupervised approaches

Machine learning methods, like those used in natural language processing and artificial intelligence, are generally divided into two types of methods: supervised and unsupervised. Supervised methods are trained on a dataset that has been analyzed or labeled ahead of time, and they aim to create a model which can predict patterns or analyze features in unseen data. In some cases, this data might simply be a collection of multiple statistics, like student demographics and time to graduation [16], while in others it might involve human coders labeling images or pieces of text. Supervised methods are commonly used across many different fields and applications, since they are straightforward to evaluate: given a labeled dataset, one can compare the predictions of the model on a subset of the data to the actual labels, and quantify their level of agreement or disagreement. However, they can be expensive to train, as creating a suitable coded/labeled dataset often requires significant investment of time, effort, and resources.

Unsupervised methods, in contrast, are trained on data where there is no a-priori labeling or classification. For example, data might be grouped together using a clustering algorithm like K-Means or a dimensional reduction algorithm like principal component analysis.[2] Unsupervised methods have the advantage that they do not require data to be labeled ahead of time in order to work; however, this has the disadvantage of making them much more difficult to interpret, evaluate, or benchmark.

When it comes to applying natural language processing techniques to qualitative analysis in PER, as well as some related work in Science Education Research (SER), we can map out the space of prior work according to the axes of deductive/inductive and supervised/unsupervised to examine the different research paradigms and contextualize the present study. This mapping is shown in Figure 1.

---

[2] For a more thorough overview of machine learning methods applied to science education, we recommend Wulff (2023) [17]

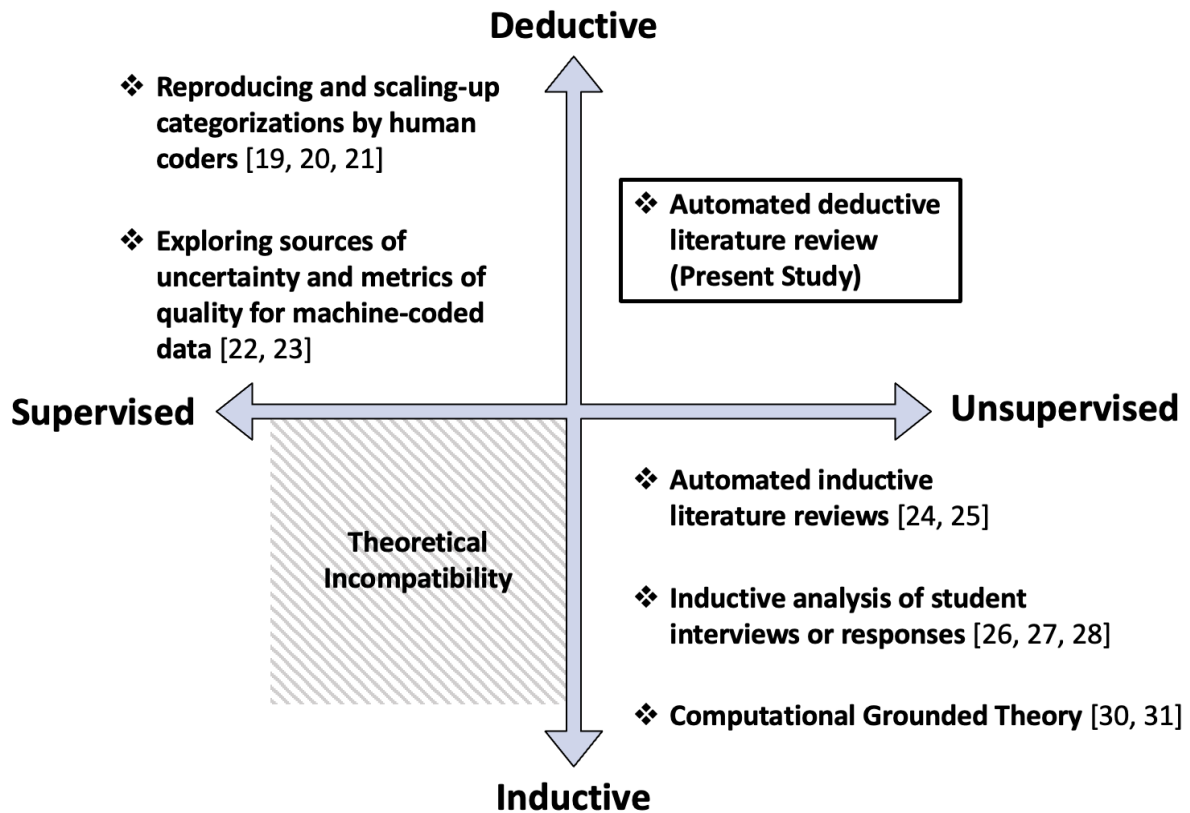

**FIG. 1.** Space of the literature in physics education research related to deductive/inductive qualitative analysis and supervised/unsupervised natural language processing.

1. ***Supervised NLP applied to deductive qualitative analysis***

This paradigm is defined by researchers creating models that learn and apply pre-existing coding and classification schemes to textual datasets using standard supervised NLP methods, like logistic regression. According to a review by Zhai et al. of machine learning methods in science education assessment (usually a deductive task) most such studies use supervised methods [18].

Within PER, much of this work has focused on creating models that can learn to reproduce classifications made by human coders, in the hopes that such models could be used as a tool for rapid analysis of arbitrarily large future datasets. For example, Wilson et al. (2022) created a logistic regression model that categorized student reasoning on the Physics Measurement Questionnaire (PMQ). It was trained using a human-coded dataset of approximately 2450 responses, categorized using a coding scheme focused on different types of reasoning about measurement. The model produced similar levels of inter-rater reliability as that between two human coders [19].

Some researchers have also explored how much data is necessary to create robust supervised deductive models. For example, Wulff et al. (2021) created a multinomial logistic regression model that classified written reflections by German pre-service physics teachers, which had again been coded by human researchers ahead of time. Despite being trained on a relatively small amount of data (93 reflections) the model was able to achieve acceptable agreement with human coders on at least some of the code categories [20]. Wulff, Mientus, et al. (2022) performed a similar study with a larger dataset (N = 270 human-coded reflections from German pre-service physics teachers) using a German pre-trained large language model that was fine-tuned to the specific task using standard supervised machine learning methods, like hold-out and k-fold cross validation. When comparing the performance of this model with other deep learning architectures, they found that the fine-tuned classifier outperformed other deep-learning architectures. [21].

Still other researchers are exploring the sources of uncertainty and metrics of quality in these types of models. Fussell et al. (2022) and Fussell et al. (2023) trained two supervised machine learning models (a neural network and logistic regression model, respectively) to apply an existing coding scheme to student responses on an open-ended survey about the trustworthiness of experimental results. Based on these trials, they propose a framework that describes the conditions and criteria necessary to consider machine coding trustworthy and reliable [22,23].

2. *Supervised NLP applied to inductive qualitative analysis*

This paradigm is defined by researchers using supervised NLP methods to create models that inductively learn or extract new themes from datasets. So far as we know, there is no research that directly fits this paradigm, as there is a fundamental incompatibility between the epistemological commitments of inductive qualitative analysis and supervised machine learning. That is, supervised machine learning requires one to approach the analysis *with* defined categories and some amount of prior analysis, while inductive qualitative analysis requires one to approach the data *without* a defined set of categories or prior analysis.

3. *Unsupervised NLP applied to inductive qualitative analysis*

This paradigm is defined by researchers using unsupervised NLP methods to extract themes or create new categories from the data that were not known a-priori. A number of studies from PER and SER fit this paradigm. For example, Odden et al. (2020) used an unsupervised topic modeling method called Latent Dirichlet Allocation (LDA) to perform a standard inductive qualitative research task: a literature review. The literature base was 18 years of the *Physics Education Research Conference (PERC) Proceedings* (approximately 1300 articles). The analysis revealed 10 distinct and recurring research topics and quantitatively mapped out their rise and fall over time. [24]. Odden et al. (2021) used LDA to perform a similar analysis on approximately a century's worth of literature from the journal *Science Education* (approximately 5570 articles), again mapping large-scale shifts in that literature. The analysis showed that the journal had undergone a shift from science content-based articles, to studies focused primarily on teaching, and then to studies focused on student learning. It also showed that many of the advances in the field happened when science education borrowed theoretical frameworks or paradigms from other related fields like psychology and sociology. [25].

Other researchers have used unsupervised methods for inductive analysis of student-level data. Clustering methods are especially popular in this area. For example, Sherin (2013) used clustering algorithms to analyze segments of an interview transcript on the topic of how the Earth's orbit affects the seasons, finding that the clusters were thematically interpretable and suggestive that such methods could be used to enhance inductive qualitative analysis [26]. Wulff et al. (2022) used a clustering approach to categorize physics pre-service teachers' written reflections (N = 86) of a teaching situation. Some amount of prior data analysis was performed using the trained classifier from Wulff, Mientus, et al. (2022), which extracted relevant sentences according to a pre-defined coding scheme. These sentences were then grouped into topics using an unsupervised clustering algorithm, HDB-SCAN. The resulting clusters were found to be interpretable, and allowed the researchers to organize and visualize different themes in the responses [27]. Geiger et al. (2022) used LDA to identify specific student ideas about circuits, based on a set of approximately 500 responses to a conceptual question about bulb brightness in a circuit with one vs. two batteries. The resulting set of five ideas were quite interpretable, focusing on concepts like analogies to water flowing through circuits, different effects of electric potential, and quantitative relationships in Ohm's law [28].

Some researchers have explicitly positioned their analyses as a specific variety of qualitative analysis, such as grounded theory, and used that choice to guide the way they combine NLP and human analysis methods. For example, Rosenberg and Krist (2021) analyzed (N = 173) responses from a 7th-grade chemistry assessment using a Computational Grounded Theory (CGT) approach, based on Nelson (2020) [29], in order to understand the students' epistemic ideas about generalizability. In this analysis, the authors first (inductively) produced an initial construct map that summarized their hypotheses about the data. Responses were then clustered using an unsupervised algorithm, and then these clusters were reviewed and compared to the construct map to produce a new coding scheme, which was again applied to the data. Finally, the authors used this coding scheme to train several supervised algorithms, which were found to have substantial agreement with human codes [30]. Tschisgale et al. (2023) also used a Computational Grounded Theory approach in order to analyze student solutions to a German Physics Olympiad problem, comparing them to a control group of students who solved the same problem but did not participate in the Physics Olympiad. Their approach used a dimensional reduction technique called UMAP, followed by the unsupervised clustering technique HDB-SCAN to categorize 1127 sentences from the dataset. The researchers then qualitatively interpreted the resulting clusters and grouped them into themes. This analysis revealed that Olympiad participants showed more expert-like problem-solving behavior (referring more to assumptions and idealizations, conceptual aspects, and quantitative aspects of problem solutions) compared to non-participants [31].

4. *Unsupervised NLP applied to deductive qualitative analysis*

This paradigm is defined by researchers analyzing a dataset by applying pre-existing categories to it using unsupervised NLP methods. Although there might appear to be another theoretical incompatibility between these two approaches, this is only the case when one is using completely unsupervised methods (i.e., those without any input from a researcher) to perform deductive analysis. In practice, most unsupervised methods require some amount of researcher input, both in initialization and in tuning of model hyperparameters; for example, topic analysis methods like LDA allow researchers to specify the number of topics and the general "mixedness" of topics, while clustering methods allow researchers to specify the number of clusters and choose (if they wish) the initial centroid of each cluster. Thus, certain unsupervised NLP methods could, in principle, be used for deductive qualitative analysis.

To our knowledge, this paradigm has yet to be explored in the PER and SER literature. This represents a significant gap, as such a paradigm could provide significant advantages in flexibility and ease of use, combining the resource efficiency of unsupervised methods (which don't require humans to label large amounts of data ahead of time) with the many applications of deductive qualitative analysis.

In this article, we show a proof of concept of this kind of analysis. To do so, we have used a cutting-edge NLP technique called text embeddings, which allows one to turn texts into high-dimensional vectors and then perform vector operations on them, like calculating distance and similarity measures.

In what follows, we describe the theory behind text embeddings and contrast them with the features of a previously-used technique, Latent Dirichlet Allocation. We then use text embeddings to replicate an inductive review of PERC Proceedings articles [24], extending that analysis up to the most recent published year as of writing (2023). This provides a benchmark for the technique and a proof of concept that it can be deductively applied to reproduce prior results. It also allows us to evaluate its affordances and drawbacks. We then extend the analysis by performing a second literature review of PERC proceedings articles using researcher-defined topics, in order to show the versatility of the method.

### C. NLP theory: LDA and embeddings

Both LDA and text embeddings rely on the fundamental idea of representing text as vectors. This representation allows researchers to quantitatively explore different aspects of the text—such as word co-occurrence, meaning, sentiment, etc.—using standard tools for vector manipulation, such as addition, subtraction, projection, and decomposition. However, there are different ways to construct these kind of representations, which hinge on the size and meaning of the vector space into which one projects the text.

Latent Dirichlet Allocation (LDA) [32,33] is based on a Bag of Words (BoW) model. This approach, which has a long history in NLP research [34] represents each text in the dataset as the count of the unique words in the document. Thus, when creating a bag of words out of a set of texts, each document will be represented as a vector, with a dimensionality equal to the total number of unique words across all the texts. Each entry in a document's vector corresponds to a particular word, and the entry's value is the number of times that word occurs in that document.[3]

From a vector perspective, a BoW model thus treats each word as an independent dimension in a high-dimensional vector space encompassing all the words in all the documents (in practice, often many thousands of dimensions). A document is then represented by a specific point (or vector) in this space. Depending on the number, length, and heterogeneity of documents, such vector spaces may be sparsely populated. But, with enough documents and enough words, statistical regularities can emerge—that is, certain documents may use similar words, putting them at similar "locations" in the high-dimensional "word space." Topic analysis techniques like LDA allow one to use these regularities to extract latent *topics*, in the form of sets of words that frequently occur together [32,33]. Each document can then be scored based on the amount (prevalence) of these topics in its text, which, in turn, allows one to look for trends in topic prevalence across the dataset. For example, if one has a large number of documents over a significant time period, one can evaluate how the prevalence of different topics has varied as a function of time [24,25,35]

However, both LDA and the underlying Bag of Words model have some limitations and implementation challenges. For starters, such analyses often require extensive data pre-processing and filtering, since words that do not help to distinguish semantic meaning (commonly known as *stop words*, like "is", "and", "but", etc.) must usually be filtered out, along with words that are over-represented in the dataset [24]. Because LDA is an unsupervised algorithm, it can also be a challenge to interpret identified topics, and even when topics are interpretable they may not necessarily reflect meaningful trends from a researcher perspective. For example, in Odden et al.'s 2020 analysis of PERC proceedings, one of the identified topics focused primarily on student difficulties with quantum mechanics. This topic likely emerged because these types of studies use a particular vocabulary that is distinct from that in most other PERC articles; however, from a researcher perspective it would have been more helpful to have a topic that focused, for example, on student conceptual difficulties across multiple physics topics. Finally, topic modeling techniques are often unstable, in models are randomly initialized and results may be sensitive to these initialization states [24].

---

[3] For example, the 0$^{\text{th}}$ entry might correspond to the word "physics", and the 1$^{\text{st}}$ entry might correspond to the word "education"; a document that uses the word "physics" 10 time and "education" 12 times would have 10 as the first entry in its vector, 12 as the second, and so forth.

Embeddings are a more sophisticated method for representing words and text, which do not treat words as independent dimensions. Instead, the words themselves are transformed, using a large language model, into high-dimensional vectors (usually several hundred dimensions) which live in a semantic space that encodes their relative meanings. This allows words to be manipulated using classic vector-handling techniques: for example, one can add and subtract word vectors, such as taking the word-vector for "king", subtracting "man", and adding "woman" to construct a vector that lies closest in the semantic space to the word "queen" [13]. One can also use dot products to evaluate their similarity (often known as a cosine similarity metric) and find the distance between two or more vectors using various distance metrics (like Euclidean distance or Manhattan distance).

Many of the recent advances in large language models and AI chatbots are based on representing words and texts using embeddings. In 2017, the foundations of these breakthroughs were laid when researchers developed a model that allowed one to encode relationships between word vectors, using a so-called *attention mechanism*, which allowed them to develop transformer-based large language models [15]. This, in turn, allowed researchers to train language models that could take in and create embeddings from longer texts, like the Bidirectional Encoder Representations from Transformers (BERT) model [36]. These text embeddings can encode not just the content or meaning of a text but also potentially deeper features like sentiment and writing style [37] and often use vectors of several hundred dimensions (commonly 512, 768, or 1536). By representing increasingly longer texts as vectors, entire documents could be compared using the vector-handling techniques described above—for example, one can compare the similarity of documents based on their relative distances in a high-dimensional space.

Embeddings therefore offer an alternative approach for comparing and extracting latent features across documents, where these vectorized representations of documents can encode more information than pure counts of word occurrence. This opens the possibility of doing analyses similar to those that can be done with a BoW model, like topic modeling.

In this study we investigate this possibility. Rather than using embeddings to do inductive qualitative analysis, where we let an algorithm group the texts together according to some metric (as with LDA or clustering), we instead explore how one could use them deductively, choosing a set of archetypal texts from the data, aggregating them, and then using the relative distance between particular texts and these aggregated examples to classify the remainder of the data in an unsupervised way. Our research questions are therefore as follows:

1. Can embeddings be used to deductively replicate prior inductive qualitative research on scientific literature like the PERC proceedings?
2. Can embeddings be used to perform alternative analyses of this literature, using fully researcher-defined categories?
3. What do these analyses tell us about the development of the literature over time?

### III. METHODS

#### A. Dataset and update to prior LDA analysis

We based the present study on a prior dataset and study performed by Odden et al. (2020), which analyzed PERC proceedings published from 2001-2018 using Latent Dirichlet Allocation. That analysis found the following ten recurring research topics:

1. **Representations:** This topic focuses on research on student understanding of and difficulties with representations and the physics content usually used to teach representational fluency, like Electricity and Magnetism.
2. **Problem Solving:** This topic focuses on research on problem solving in physics.
3. **Laboratory instruction:** This topic focuses on research on physics laboratory instruction.
4. **Assessment (Concepts):** This topic focuses on use of quantitative methods to measure student conceptual understanding, such as concept inventories.
5. **K-12:** This topic focuses on research on pre-college physics education, such as teacher professional development and pedagogy.
6. **Quantum Difficulties:** This topic focuses on student understanding of and difficulties with quantum mechanics, although the topic also shows up in articles related to student difficulties in other subjects like E&M.
7. **Qualitative Theory:** This topic focuses on theoretical case studies of physics student cognition, usually qualitative in nature, with a specific focus on student discourse and explanation.
8. **RBIS:** This topic focuses on studies related to research-based instructional strategies, such as tutorials and clickers, in lectures and recitation sections.
9. **Assessment (Demographics):** This topic focuses on quantitative research that unpacks the effects of different factors in student learning, such as gender and ethnicity.
10. **Community and Identity:** This topic focuses on research that uses more sociocultural perspectives, such as communities of practice, institutional change, and physics identity.

The analysis also quantified the prevalence of these topics in the data corpus of approximately 1300 PERC proceedings articles published from 2001-2018, by scoring each article based on the percent of each of these 10 topics in the article's text. Thus, one article might for example include 50% RBIS, 20% laboratory Instruction, and 30% Problem Solving topics, while another article might contain 80% Community and Identity, 20% Assessment (Demographics).

In line with our research questions, our first goal was to explore whether embeddings could be used to deductively reproduce this prior NLP-based inductive literature review of PERC proceedings. To explore this question, the remaining PERC literature published 2019-2023 was downloaded and the text was extracted using the same text extraction tools as in the 2020 analysis. This resulted in a data corpus of 1745 articles. Next, an identical data cleaning procedure was used to that from Odden et al. (2020) to produce an up-to-date dataset for LDA analysis. Using the LDA model created by Odden et al. (2020), we scored the new articles based on these 10 topics (using built-in functionality from the Gensim implementation of LDA). We present the results of this analysis in Part A of our Results; however, the primary purpose of this analysis was to provide a set of scores to benchmark our deductive analysis using embeddings.

### B. Description, implementation, and benchmarking of method

The method for this study consisted of four primary steps: 1) create embeddings vectors out of texts; 2) define topic centroids based on a selected sample of texts that exemplify desired features; 3) calculate and transform distances to produce topic scores, using scaling parameters to determine "mixedness" of the model; and 4) evaluate results of the analysis based on face validity of topic scores and general trends. We visualize these steps in Figure 2, and elaborate on each below:

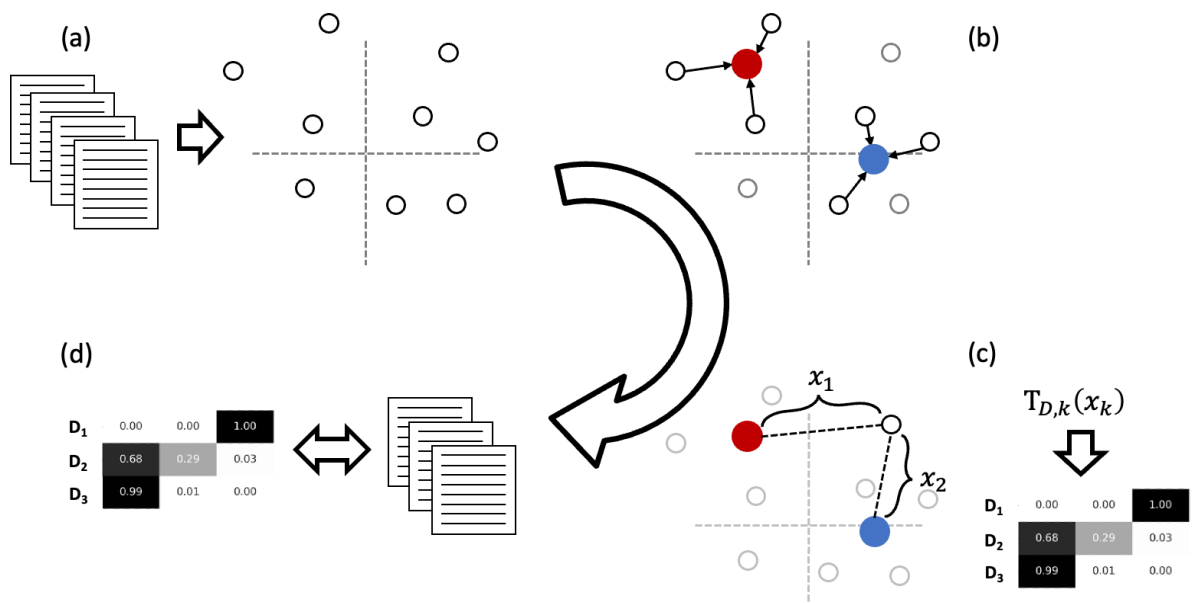

**FIG. 2.** Flowchart diagram of method for performing deductive qualitative analysis using embeddings: (a) Creating embeddings vectors from texts; (b) Defining topic centroids based on a sample of texts featuring that topic; (c) Calculating distances from texts to topic centroids, and transforming these into topic scores using a transformation function; (d) Evaluate results of the analysis based on face validity of topic scores and general trends.

#### 1. *Creating embeddings vectors out of texts*

To begin, we created embeddings out of the raw text of each article, using an open-source large-language model called *Jina Embeddings 2* [38]. This model was chosen because it is one of the few open-source models which allows one to create embeddings out of texts of up to 8192 tokens (individual words, numbers, etc.) in length. It can be run locally using an implementation from the HuggingFace repository[4]. The large language model took in the text of each article, and produced a 512-dimensional vector with vector entries varying between -1 and 1 which represented the meaning of the text in the 512-dimensional semantic space.

### 2. *Defining topic centroids based on a selected sample of texts that exemplify desired features*

Next, we defined specific locations in the semantic space that corresponded to particular topics. To do so, for each of the 10 topics from the LDA analysis, we selected the 15 most representative articles (that is, the articles with the highest percentage score of that topic, often 90% or more) and took the mean of their embeddings vectors. This produced 10 mean embeddings, one for each topic, which we refer to as the *topic centroids*. Since these topics were meant to capture similar meaning to the LDA topics, we used the same topic names and descriptors.

### 3. *Calculating and transforming vector differences to produce topic scores*

Next, we aimed to classify each article based on the relative distance between it and the various topic centroids in the semantic space. To do so, we used a similarity-based approach [39], similar in some ways to the technique known as semantic search [40], in which we first computed the distance between every individual article in the corpus and each topic centroid. For the purposes of exploration, we computed these distances multiple times using several distance metrics: Euclidean distance, cosine distance (1 - dot product), and Cityblock (Manhattan distance). Due to the high dimensionality of the space, these distances were typically very similar (for example, dot product scores of 0.91-0.96, or equivalent cosine distances of 0.04-0.09) which yielded small-but-meaningful differences in distance from every paper to each topic centroid. So, we applied a transformation function to the distances in order to invert them (since smaller distances should produce larger topic scores) and scale them up, after which we normalized these scaled distances for each document to produce a set of topic scores that summed to 1 with each individual score ranging from 0 to 1. We tried out two different transformation functions:

$$T(x) = e^{-\alpha x} \quad (1)$$
$$T(x) = x^{-\alpha} \quad (2)$$

Here, α is a hyperparameter which in practical terms controls the degree of homogeneity/heterogeneity of the resulting topic distributions, similar to the α hyperparameter in LDA [32].

Embeddings model results were then benchmarked and evaluated by computing and summing the Jensen-Shannon divergence (JS divergence) between the embeddings topic scores and the LDA topic scores across the entire data corpus. JS divergence is a measure often used in data science applications to quantify the differences between two distributions, based on methods from the field of information theory. A value of 0 indicates that the two distributions are identical, and increasing amounts of difference are represented with increasingly higher divergence values. In our case, the distributions in question were the percent topic scores by LDA compared to same topic scores by the embeddings model for each paper (illustrated in Figures 6 and 7); we thus computed the JS divergence on each article, and summed the results to get an aggregate divergence for the complete data corpus.

---

[4] This avoids issues with other proprietary embeddings models, such as those from OpenAI and Google, which require one to upload articles to their systems using an API.

After running approximately 500 models to evaluate all combinations of the two transformation functions and three distance metrics over a range of α values (0-200), we found that the Jensen-Shannon divergence was minimized when we used the normalized exponential function (Eq. 1) with approximately equal minimum values for cosine distance (α = 94) and Euclidean distance (α = 34), as shown in Figure 3. We therefore chose to use the cosine distance metric, since it often performs better in high-dimensional datasets [41] and is recommended by OpenAI for use in their embeddings models [42].

For a given document D and topic $k \in \{1, …, K\}$ we therefore define the normalized topic score from the embeddings model, $T_{D,k}$, as:

$$T_{D,k}(x_k) = \frac{1}{\sum_{k=1}^{K} e^{-\alpha x_k}} e^{-\alpha x_k} \quad (3)$$

Where $x_k$ is the cosine distance between the document and the $k^{th}$ topic centroid, calculated by taking 1 minus the dot product between the vectors; and α is a hyperparameter controlling the relative "mixedness" of the topics. This score can take a value of 0-1 (corresponding to the percentage of the $k^{th}$ topic in document D), and the scores of all topics k = 1…K for each document sum to 1. For reference, this divergence was also compared to that produced by uniform data and the mean of 2000 runs of random data. The minimum JS divergence was found at α = 94, as shown in Figure 3.

As a second measure of divergence between the two models, we also calculated the summed absolute value of the topic-by-topic difference between LDA and embeddings scores for each article. For a particular article, this value could hypothetically range from 0 (if both the LDA and embeddings models assigned identical topic scores) to 2 (if there was no overlap between topic scores), meaning the aggregate score could range from 0 to 3490 for the entire corpus. At the chosen α = 94, the calculated aggregate difference was approximately 1061, indicating that the embeddings model seems to have recovered approximately 70% (2429/3490) of the topic scores by LDA. This measure, when plotted as a function of α, also had a minimum at approximately the same α value as the JS-divergence plot.

Finally, to unpack how the two models differed in their scores of specific topics, we calculated the correlation between LDA and embeddings model scores for each topic using Kendall's Tau (computed using the Scipy stats package), an appropriate metric for data that does not pass the test of normality and which includes many ties when data is ranked (both of which were the case for this dataset). This analysis can be found in the supplemental material of the article.

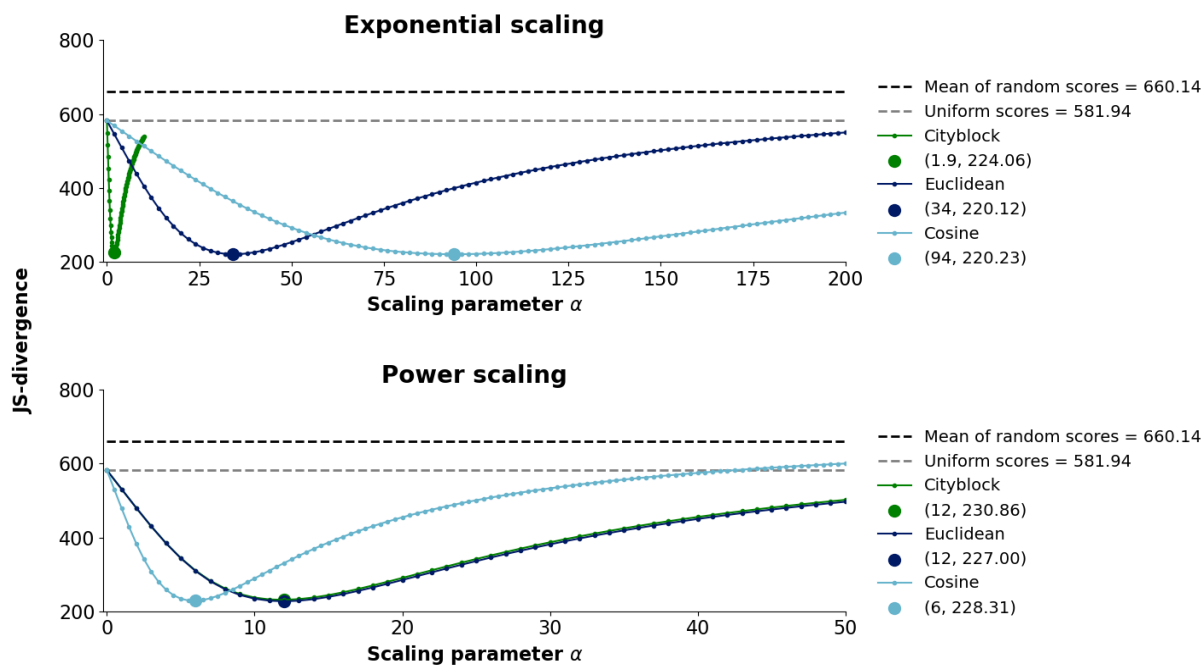

**FIG. 3**. Jensen-Shannon Divergence between LDA topics and LDA-derived embeddings topics, using exponential and power scaling, as a function of scaling parameter α. Distances were calculated using three metrics: cityblock (Manhattan distance), Euclidean, and cosine distance (1 – dot product). JS Divergence was also calculated with uniform topic distributions and random data (mean of 2000 runs) for reference. Minimum divergence is found with cosine distance metric, using exponential scaling function, at α = 94.

### 4. Evaluating results of the analysis based on face validity of topic scores and general trends.

Finally, the first author checked the face validity of the model by comparing the scores of a representative set of individual articles by LDA and the embeddings model. This embeddings model was then used to analyze the topic prevalence in PERC proceedings as a function of time, by aggregating and normalizing the percent of each topic from all articles in each year. Results were then qualitatively compared with those from the updated LDA analysis, and differences were unpacked in light of the different textual features incorporated by the two methods.

### C. Extension of the analysis: embeddings-based literature review of PERC Proceedings based on Docktor & Mestre (2014)

Having used the LDA analysis as a benchmark to explore the use of embeddings for qualitative analysis, we next set out to perform a new analysis, using fully researcher-defined topics based on the PER literature review by Docktor & Mestre (2014). To do so, we first re-downloaded the entire dataset and scraped the text using a more up-to-date text extraction library, which provided better quality texts sensitive to the two-column format of most PERC proceedings articles. This allowed us to retain several articles that had been removed from the prior analysis due to text extraction issues, bringing to complete corpus to 1761 articles.

Next, we repeated all of the methodological steps described above, except for benchmarking of results against prior analyses. That is, we first created new embeddings vectors of all 1761 articles using the same *Jina Embeddings 2* large language model described above [38]. We then defined a new set of topic centroids by operationalizing the categories and sub-categories proposed by Docktor & Mestre (2014) in their review and synthesis of physics education research:

1. **Conceptual understanding:** Including research on *Naïve theories or misconceptions*; *Knowledge in pieces or conceptual resources*; and *Ontological categories or metaphors.*
2. **Problem Solving:** including research on *Expert-novice differences; Worked examples; Representations; Mathematics in problem solving;* and *Instructional strategies for problem solving.*
3. **Curriculum and Instruction:** Including research on *Lecture-based methods; Recitations and discussion-based methods; Laboratory instruction; Structural changes to classroom environments;* and *General instructional strategies and materials.*
4. **Assessment:** Including research on *Development of concept inventories; Comparing scores across multiple measures; Comparing scores across multiple populations and demographics; Course exams and homework;* and *Rubrics for assessment*[5].
5. **Cognitive Psychology**: Including research on *Knowledge and memory; Attention; Reasoning and problem solving;* and *Learning and transfer.*
6. **Attitudes and Beliefs:** *Including research on Student attitudes and beliefs; Faculty attitudes and beliefs;* and *Teaching assistant and learning assistant attitudes and beliefs*[6].

On top of these categories, we added two additional categories encompassing literature not included in that review, either due to explicit exclusion criteria (7) or because it had not yet emerged as a major theme in the literature base (8):

7. ***Pre-College Physics Education:*** Including research on *Teaching tools; Teacher preparation; Teaching strategies;* and *Student experiences.*
8. **Identity and Equity:** Including research on *Physics identity; Equity; Race;* and *Gender.*

To define new topic centroids, the first author selected 4 representative articles for each topic and sub-topic from a list of all articles in the data corpus, aiming for a variety of authors, years, and study topics in each sub-cluster (the complete list of chosen articles can be found in the supplementary material). Articles were chosen based on whether they referenced some aspect of the sub-topic in their title: for example, using the terminology "expert/novice" in a study of problem solving, or "physics identity" in a study on identity and equity. Thereafter, we constructed centroid vectors from these articles by taking the mean of their embeddings vectors, and computed topic scores using the same distance and transformation functions as before (Eq. 3) with a high alpha value ($\alpha = 200$) to encourage the model to assign high primary topic scores.

Inherent to this technique is a sensitivity to the particular articles used to define the centroids. To test this sensitivity, we conducted a centroid resampling procedure in which we created new topic centroids by repeatedly sampling the sub-topic articles from each topic. Specifically, for each of the 8 topics, we sampled sets of 3 articles from the 4 in each sub-topic and took the collective mean of their embeddings vectors to define alternative topic centroids, which were then used to compute topic scores using the method described above. This procedure was repeated 1000 times to provide a measure of the spread of resulting topic scores. Upon investigation, the mean topic scores from these 1000 models was most aligned with the set of topic centroids based on all available text data (that is, averaging all 4 article vectors from every sub-topic). We therefore used topic centroids constructed from all articles in teach topic to produce a final analysis of the topical prevalence as a function of time in the data corpus. The standard deviation of the 1000 resampling models was used to set error bars on this prevalence. We also produced plots of average prevalence over time, and heatmaps of topic scores for individual articles to establish model face-validity.

---

[5]In topic 4, the sub-topic *Complex Models of Student Learning* was dropped due to difficulty in operationalizing it

[6]In topic 6 the sub-topic *Instructor Implementations of Reformed Curricula* was dropped due to thematic inconsistency (i.e., this topic seemed more appropriate with Curriculum and Instruction)

## III. RESULTS

**A. Updated LDA analysis and Embeddings analysis using LDA-derived topics and centroids**

As discussed, our first methodological step was to extend the LDA analysis up to the most recent published year (2023, as of writing) and to compare the performance of an embeddings analysis, applied deductively using topic centroids based on LDA model output, to these results. Figures 4 and 5 show stacked area plots for the updated LDA model and the embeddings model, respectively. In both cases, topic prevalence has been normalized each year to account for the changing number of publications as a function of time. This normalized topic prevalence thus represents the percent of each topic of the total literature in that year.

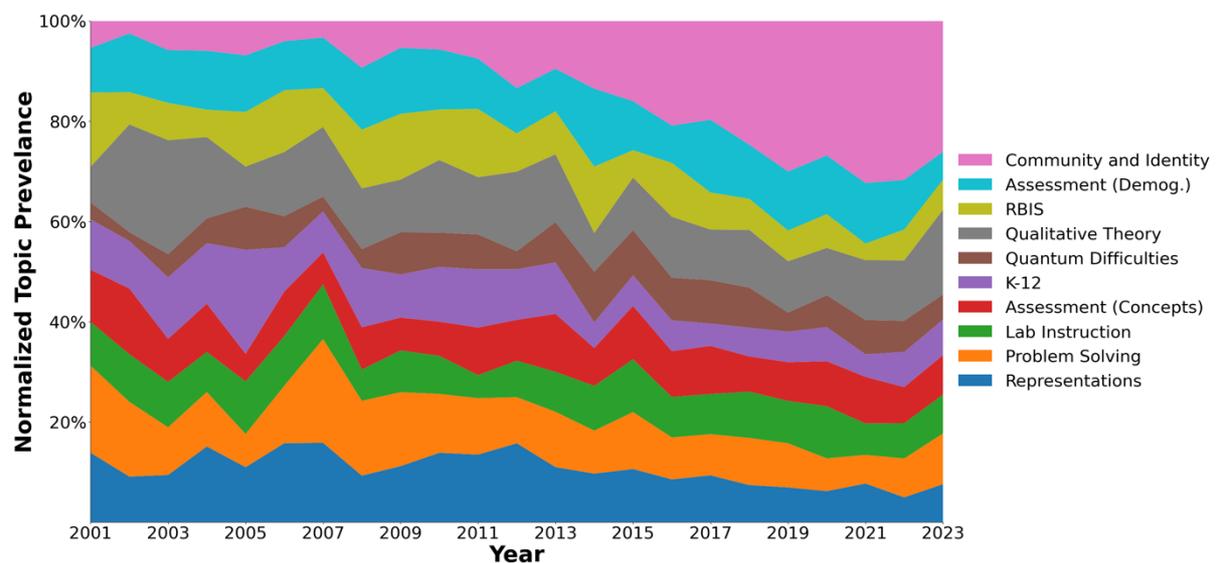

**Figure 4:** LDA-based calculation of normalized topic prevalence as a function time in PERC Proceedings 2001-2023, based on LDA model from Odden et al. (2020) [23].

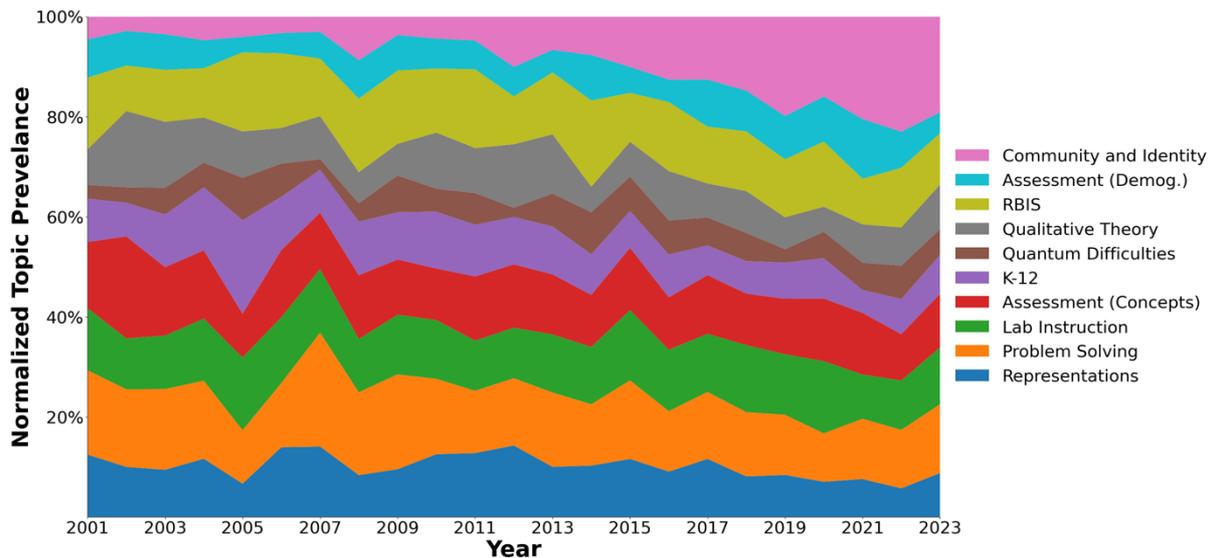

**Figure 5:** Embeddings-based calculation of normalized topic prevalence as a function time in PERC Proceedings 2001-2023, using LDA-derived centroids.

Inspection of the two plots shows that they are very similar, which indicates that embeddings model recovers a significant amount of the aggregate trends found by LDA: for example, the significant growth in the Community and Identity topic over time, a small early bump in Qualitative Theory studies, and surge in Problem Solving research in the middle years, as well as the fluctuation of research on representations. Additionally, many of the smaller spikes and dips in topic prevalence from the LDA plot are present in the embeddings model results. These similarities are to be expected, since the embeddings model has recovered approximately 70% of the total topic scores assigned by LDA.

However, there are some large-scale differences between the two models. The embeddings analysis seems to show a higher prevalence of research related to Lab Instruction, Problem Solving, and Assessment (Concepts) topics compared to LDA, and a lower overall prevalence of research on Representations, Community and Identity, Assessment (Demographics), and Qualitative Theory.

To understand why the models differ, in Figure 6 we present a snapshot of articles with approximately mean JS-Divergence (mean JS Divergence = 0.14, standard deviation 0.09) between the topic scores from the two models. The articles can be considered representative of how the two models scored articles in the data corpus on average. As one can see, the topic scores by LDA and embeddings overlap significantly in many cases, which explains why the general features of the two graphs are so similar. However, the details of the topic scores sometimes differ. For example, article 6(a) *Studying Expert Practices to Create Learning Goals for Electronics Labs* was scored by both LDA and embeddings as approximately 43-48% Lab Instruction, but LDA additionally scored it as 35% Community and Identity while the embeddings model scored it as 28% RBIS. Both of these choices seem to make sense, as the article focuses on expert practice (a community of practice perspective) and learning goals-driven design. Article 6(c), *Observing Teaching Assistant Differences in Tutorials and Inquiry-Based Labs,* was scored by LDA as 77% RBIS and 15% Lab Instruction, while the embeddings model scored it as 44% RBIS and 26% Lab Instruction, with an additional 11% Problem-Solving (likely due to the focus on tutorials).

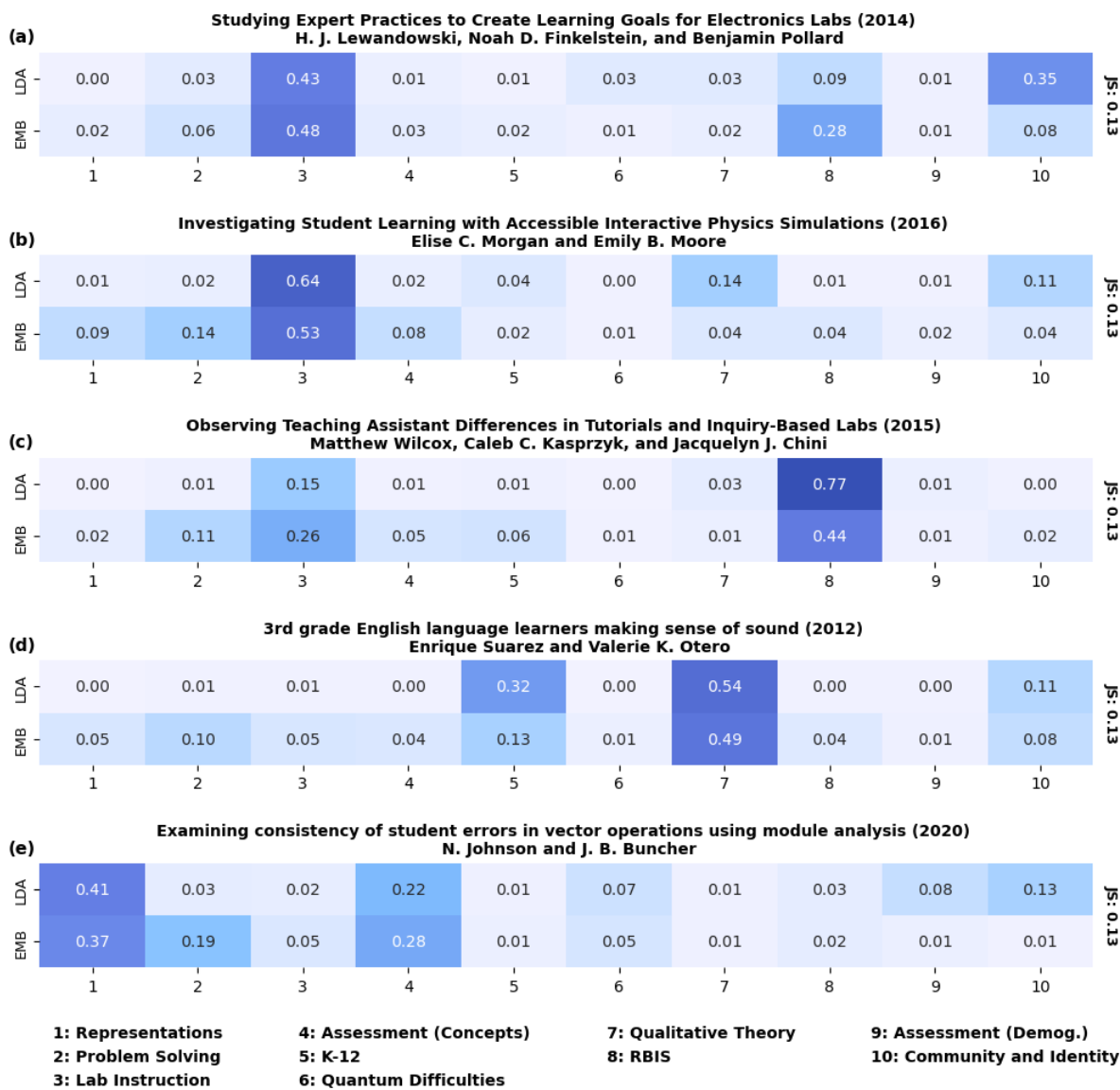

**FIG. 6.** Heatmap of topic scores for 5 articles (a-e), scored using 10 topics by LDA and Embeddings models. Articles were chosen because they have approximately mean Jensen-Shannon Divergence (measuring difference between scores) between LDA and Embeddings topic scores. Darker colors indicate higher percent of individual topics.

Figure 6 also reveals that the primary difference between the models seems often to come from cases in which one model has smeared topic scores over several topics, giving low scores across many categories. For example, both models have similar primary topic scores for article 6(d), *3rd grade English language learners making sense of sound:* 49-54% Qualitative theory, likely due to the focus in the study on student mechanistic reasoning and discourse. However, LDA has additionally scored this paper as 32% K-12 and 11% Community and Identity, while the embeddings analysis has scored it as 13% K-12 and assigned low scores over most of the remaining categories. A similar dynamic is visible in article 6(e), *Examining consistency of student errors in vector operations using module analysis*, which was scored by both LDA and the embeddings model as 37-41% Representations and 22-28% Assessment (Concepts) (likely because the study analyzed student multiple-choice responses); however, LDA has smeared the remaining topics scores over Quantum Difficulties (likely due to the focus on vectors), Assessment (Demographics), and Community and Identity, while the embeddings model has assigned a 19% score of Problem Solving.

Examination of the articles with high JS-divergence, shown in Figure 7, shows a similar dynamic in which some discrepancies can be traced to interpretable differences in model focus and others seem to be due to models smearing scores across multiple categories. For example, LDA classified four of the five articles, 7(a) *Students' Understanding of the Concepts of Vector Components and Vector Products,* 7(b) *Student difficulties with unit vectors and scalar multiplication of a vector,* 7(d) *Quantum Mechanics Students' Understanding of Normalization,* and 7(e) *Students understanding of dot products as a projection in no-context, work and electric flux problems* as primarily Quantum Difficulties (likely because the vocabulary of these articles frequently used words related to student conceptual difficulties and vectors, both of which are characteristic of this topic). The embeddings model primarily classified all four articles as focusing on Representations, which also makes sense given the focus on vectors and normalization; this is likely because the embeddings model has more sensitivity to the semantic meaning of the text as a whole. Interestingly, three of these articles come from the same research group, suggesting that the models have consistent disagreement on articles about similar topics by similar authors. In the final case, LDA has scored the article primarily as one topic—Assessment (Demographics)—while the embeddings model has smeared the topic scores over 4-5 different topics.

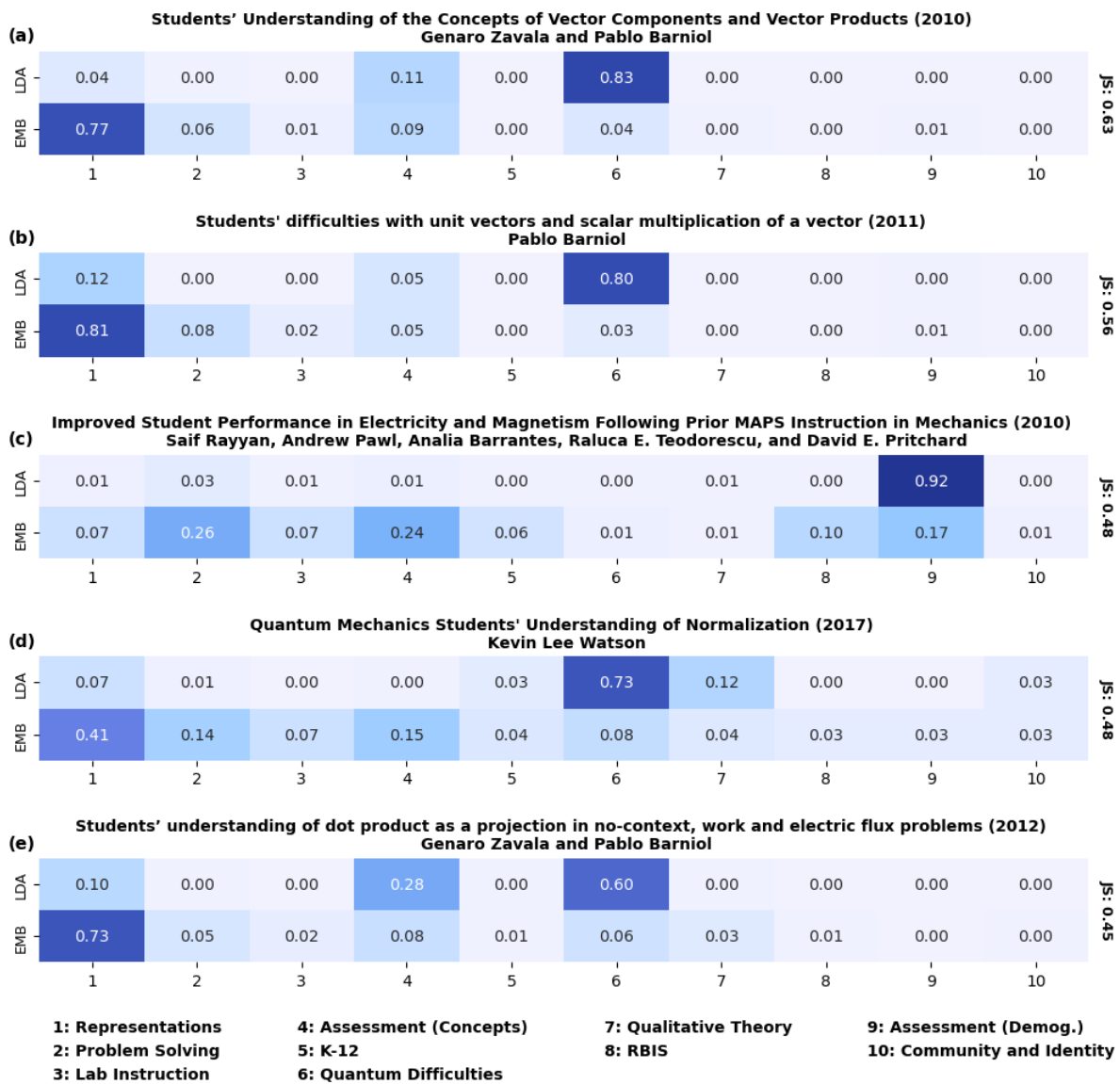

**FIG. 7.** Heatmap of topic scores for 5 articles (a-e), scored using 10 topics by LDA and Embeddings models. Articles were chosen because they had the highest Jensen-Shannon Divergence (that is, topic disagreement) between LDA and Embeddings topic scores. Darker colors indicate higher percent of individual topics.

To get a sense for how representative these differences are across the two models, Figure 8 shows histograms of the count of primary, secondary, and tertiary topic scores—that is, the highest, second highest, and third highest topic scores on each paper—from both the LDA and embeddings analyses. This histogram shows that at the chosen α value (94), the embeddings analysis tended to produce more mixed topic scores than LDA, with a slightly lower mean primary topic score (mean 0.47, standard deviation 0.20) and a peak around 0.3. In contrast, the distribution for LDA's assigned primary topic scores was higher (mean 0.53, standard deviation 0.17) and its primary topic score distribution peaked around 0.45. This indicates that LDA had a tendency to produce more distinct and focused topic scores than the embeddings model (at the chosen alpha value). We also note that two of the articles in Figures 5 and 6, articles 5(c) and 6(c), were in fact used to define the centroids for the embeddings analysis. This shows that even articles used to define centroids can still receive mixed topic scores from the embeddings model.

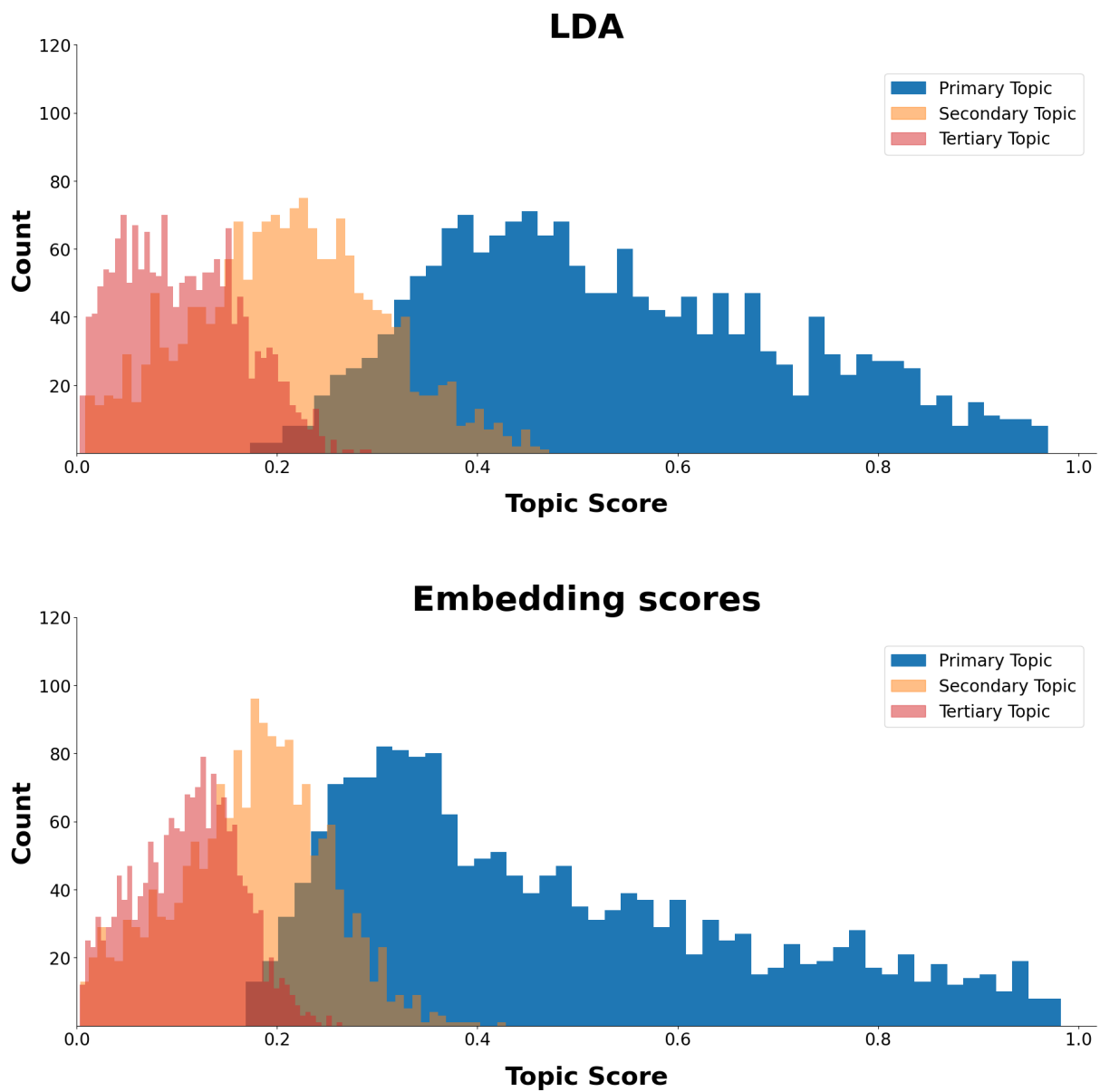

**FIG. 8.** Histogram of primary, secondary, and tertiary topic scores for each article, classified by LDA and LDA-derived embeddings with α = 94.

What do these differences mean for our interpretation of the different models? First, we stress that this tendency in the embeddings model is due in large part to our choice of α scaling parameter. Exploratory modeling showed that with increasing α, the mean of the primary topic distribution for the embeddings model shifted toward 1 (as is shown in Figure 10), and significantly more articles were classified as nearly 100% of a single topic. Furthermore, some differences are to be expected, as the two models are attending to different features of the text and arriving at their topic scores using very different mathematical models: LDA is attending to vocabulary use in the texts, and essentially performing a kind of probabilistic matrix factorization in order to extract a small number of "basis vectors" for this vocabulary [33]. Embeddings, in contrast, focus on the meanings of the texts (operationalized as aggregate relationships between the words) and are providing topic scores based on text similarities in a high-dimensional "meaning space". Thus, it seems natural to expect that the two models will often differ in their particulars, much as two human coders might differ when attending to slightly different features of a dataset. Even given these differences, the two models have a high degree of overlap (approximately 70%), both in aggregate features (Figures 4-5) and categorization of specific articles.

It should also be stressed that, prior to the current study, it was not clear that a tool like embeddings, applied deductively using a small sample of articles (15 per topic), would be able to reproduce *any* significant features of a topic analysis like that produced by LDA. Thus, this result also acts as a proof-of-concept that such an analysis is possible, as well as offering initial ideas on how to benchmark embeddings-based results to other human or machine-derived analyses. Although there are clear differences in results, we argue that these differences should be used for comparison of the two models and understanding of the embeddings analysis technique, but they do not speak to any specific "ground truth" regarding either model [35]. What can be said is that both analysis methods agree on some general trends in the data: for instance, a significant increase in studies focusing on student identities and communities of practice since around 2011; an early focus in the field on qualitative theory building; and a focus in middle years on research related to problem-solving; along with an ongoing focus on all other topics (to varying degrees).

## B. Embeddings Analysis using human-derived topics and centroids

We now turn to our second analysis, in which we used the same embeddings-based approach as before to again analyze the PERC proceedings literature, but this time used researcher-defined categories based on the review and synthesis of Discipline-Based Education Research in physics by Docktor and Mestre (2014), plus the additional categories of *Pre-College Physics Education* and *Identity and Equity*.

We first examine the face-validity of this model in order to establish whether it is, in fact, producing interpretable results. To begin, Figure 9 shows a heat map of the individual articles with highest topic prevalence (not used to define the topic centroid) for each of the 8 topics. These articles can be considered characteristic of each topic, and an inspection of these articles confirms that they all are substantively related to their primary topic, often referencing named research methods, pedagogical strategies, or theoretical frameworks in their titles. For example, article 9(f), *Gender bias in the force concept inventory?,* was classified as 100% Assessment; this makes sense, given the reference to the Force Concept Inventory. Article 9(e), *An Activity-based Curriculum for Large Introductory Physics Classes: The SCALE-UP Project,* was classified at 96% Curriculum and Instruction, which fits well considering the fact that it focuses on the well-established SCALE-UP pedagogy. Article 9(c), *Bridging Cognitive and Neural Aspects of Classroom Learning,* was classified as 94% Cognitive Psychology, which also makes sense considering the study's focus on cognition and neuroscience. The other articles show similar levels of interpretability, solidly matching specific sub-topics from the literature review by Docktor and Mestre.

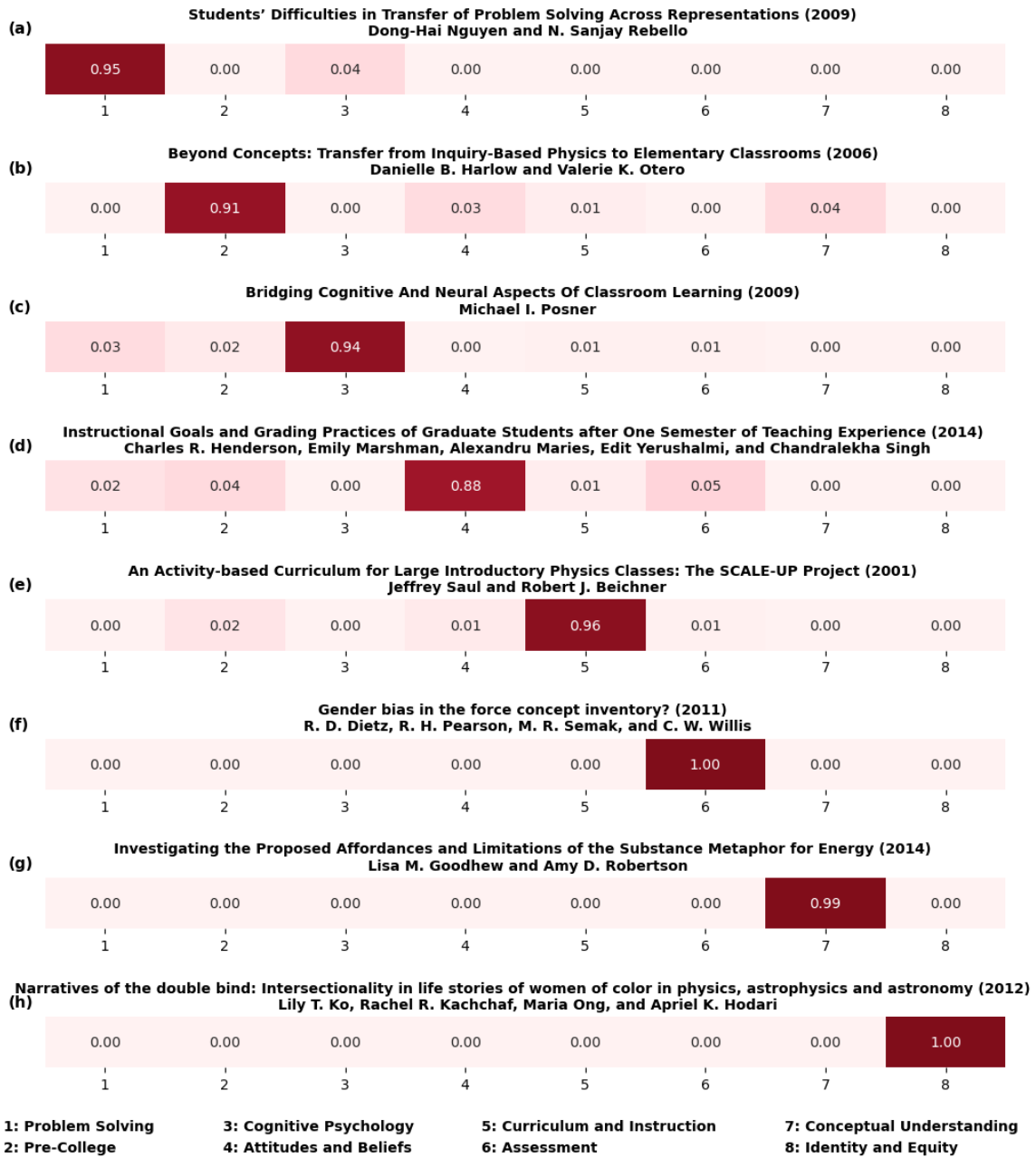

Heatmap of embeddings model topic scores for 8 articles (a-h), scored according to 8 researcher-defined topics. Articles were chosen because they have the highest prevalence of each researcher-defined topic but were not included in centroid definitions. Darker colors indicate higher percent of individual topics.

Importantly, this embeddings-based analysis is still a mixed-topic model, as shown in the histogram of Figure 10. Even though the scaling factor alpha has been set to a high level ($\alpha = 200$), the distribution of primary topics for articles varies between 0.2 and 1, with a mean of approximately 0.58 and standard deviation 0.2. Although a substantial number of articles have been classified as approximately 100% one topic (as can be seen by the peak at Topic Score = 1 in the primary topic distribution of Figure 10), most have a mixture of topics. So, in Figure 11, we present a sample of articles whose primary topic has a score of around 50%, as a second face validity check.

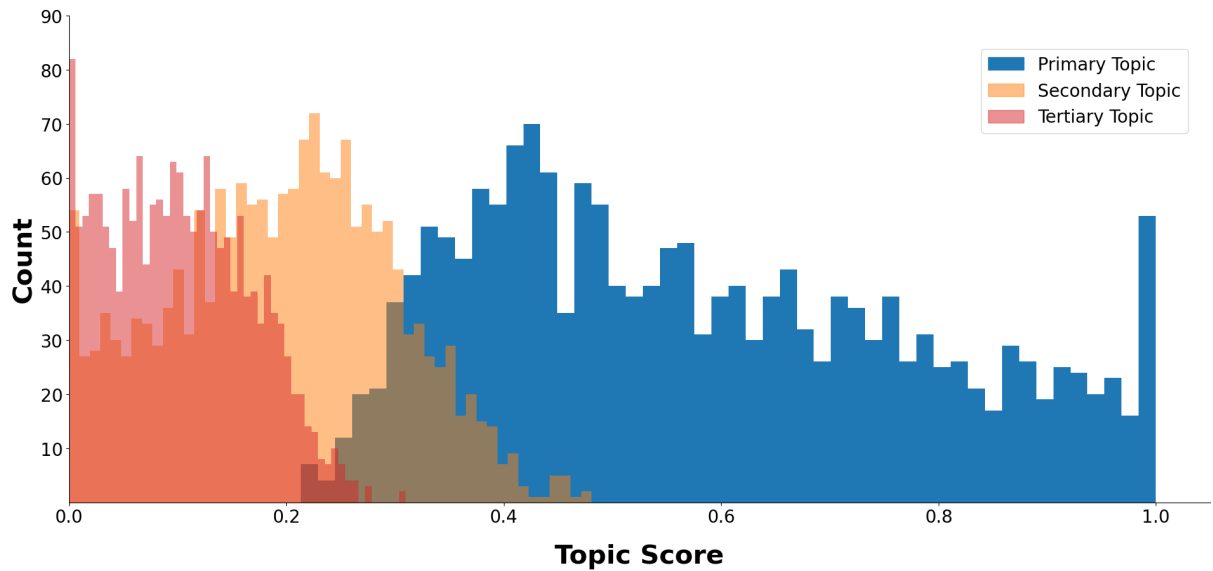

**FIG. 10.** Histogram of topic scores for primary, secondary, and tertiary topic socres for each PERC proceedings article. Scores were produced using researcher-defined topic centroids with embeddings model, α = 200.

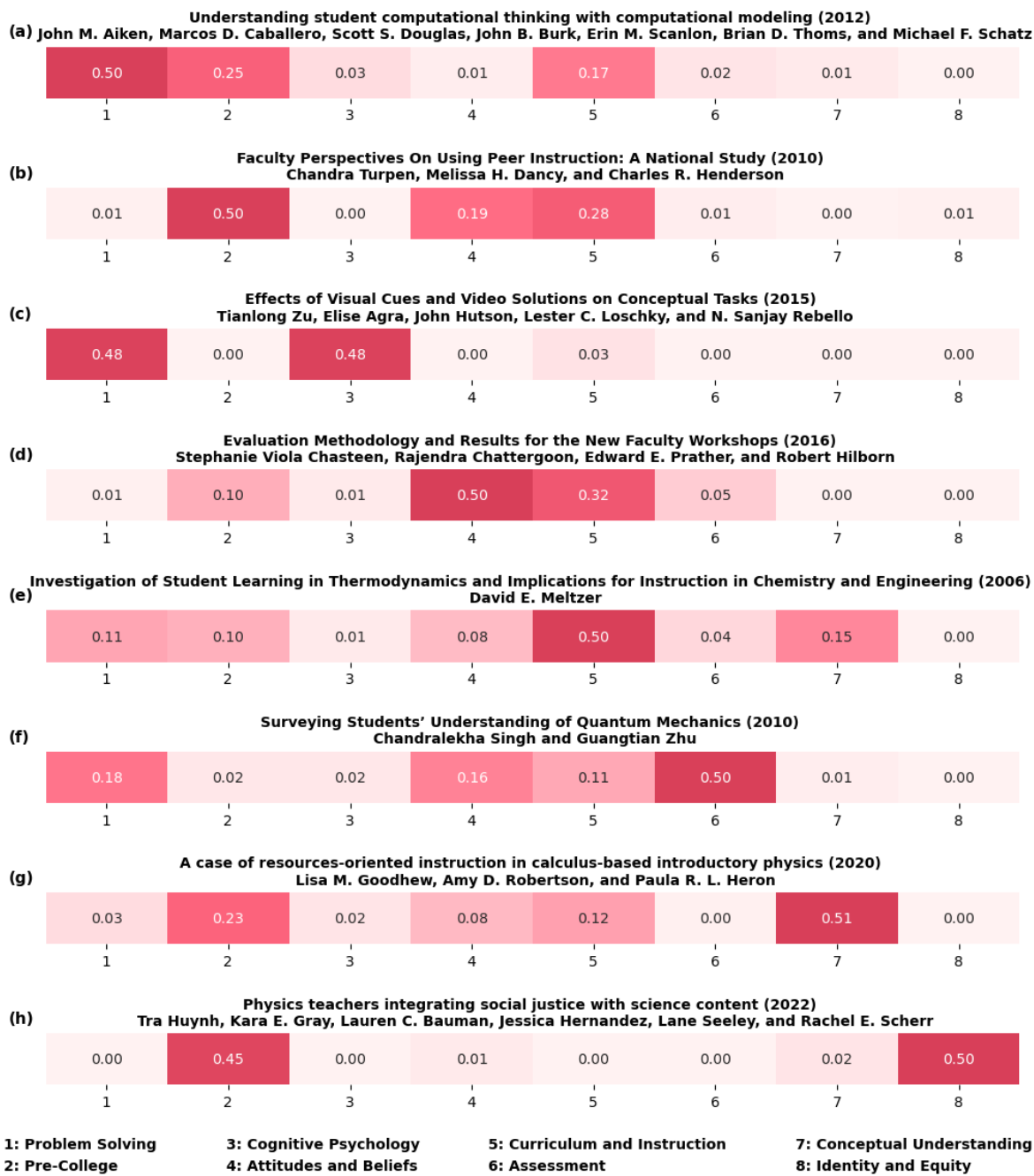

**FIG. 11.** Heatmap of embeddings model topic scores for 8 example articles (a-h) scored according to 8 researcher-defined topics. Articles were chosen because each includes approximately 50% prevalence of a researcher-defined topic. Darker colors indicate higher percent of individual topics.

Figure 11 shows that articles with a mixture of topics are also usually quite interpretable. For example, article 11(a), *Understanding student computational thinking with computational modeling,* was rated as 50% Problem-Solving, 25% Pre-College Physics Education, and 17% Curriculum and Instruction. This article analyzed how 9th grade students learned computational thinking practices while taking a variation of Arizona State University's Modeling instruction curriculum, assessed using a proctored programming assignment; it therefore makes sense that the article would be scored as a combination of problem-solving (due to the programming assignment), pre-college physics education (due to the 9th-grade students) and curriculum and instruction (due to the modeling curriculum). Article 11(c), *Effects of Visual Cues and Video Solutions on Conceptual Tasks,* analyzed how students solved physics transfer tasks using different kinds of video solutions; it thus makes sense that this article was classified as 48% Cognitive Science (due to the transfer task focus) and 48% Problem Solving (due to the worked solutions). Article 11(g), *A case of resources-oriented instruction in calculus-based introductory physics*, was scored as 51% conceptual understanding, 12% Curriculum and Instruction, and 23% Pre-College Physics Education. This study focused on a curricular approach based on the conceptual resources model, and the authors had an explicit focus on TA professional development based on literature from K-12 teacher training, which explains these scores. Article 11(h), *Physics teachers integrating social justice with science content,* was rated as 50% Equity and Identity, 45% Pre-College Physics Education, a reasonable score for a paper about high school physics teachers making connections between their teaching and social justice.

Figure 11 also shows, however, that some papers were given low scores across several categories. For example, article 11(e), *Investigation of Student Learning in Thermodynamics and Implications for Instruction in Chemistry and Engineering,* received scores of around 10-15% on four different categories (Attitudes and Beliefs, Conceptual Understanding, Pre-College, and Problem Solving). This type of spread in topic scores is a clear challenge to the interpretability this model; however, it should be noted that studies like this one may be difficult for even a human researcher to clearly classify into a single primary category, since they lie outside the standard range of PER topics and thus are not necessarily a clear fit with any single category. Such "edge cases" are common in all forms of qualitative data analysis, and so the fact that this model had difficulty classifying them is somewhat expected. We, as researchers, also have some control of the degree of topic "smear" using the tunable alpha parameter, since the number of low topic scores per article is inversely related to α (as can be seen by comparing the embeddings score distributions in Figures 8 and 10, with α = 94 and 200 respectively); however, this example shows that even very high alpha values will not necessarily eliminate such cases, especially for texts that fall outside the standard range of PER literature.

In summary, based on both inspection of specific articles and the general distribution of topic scores, we argue that the model seems to be producing results with face validity. Based on this argument, we now use this topic analysis to re-examine the development of the PERC proceedings literature over time.

Figure 12 shows plots of the cumulative prevalence over time of these 8 researcher-defined topics. Here, *cumulative prevalence* is the sum, each year, of the topic contributions from each individual article. Thus, this measure can be thought of as the approximate number of whole papers published on that topic each year. Error bars have been created using the centroid resampling procedure described above, where the width corresponds to three standard deviations of the spread from 1000 centroid resampling runs.

From this plot, certain trends stand out. First, although all topics have seen both growth in the early years (because the number of published articles has increased from around 30 in 2001-2002 to over 110 around 2019) and decline after the COVID-19 pandemic, some topics have seen significantly more growth than others. Chief among these is Identity and Equity, which grew from nearly 0 to one of the highest-prevalence topics in recent years.

Other topics have seen consistently high or low prevalence over time. Cognitive Psychology, for example, has often hovered around 5 articles per year, and Attitudes and Beliefs has had around 10 articles per year at its peak, while Curriculum and Instruction and Problem Solving have both had consistently high prevalence over most of the PERC proceedings history.

Still other topics have seen significant shifts over time. Research on Conceptual Understanding appears to have had a major surge around 2012, followed by a major dip, which is also present to a lesser degree in research on Pre-College Physics Education and Problem-Solving.

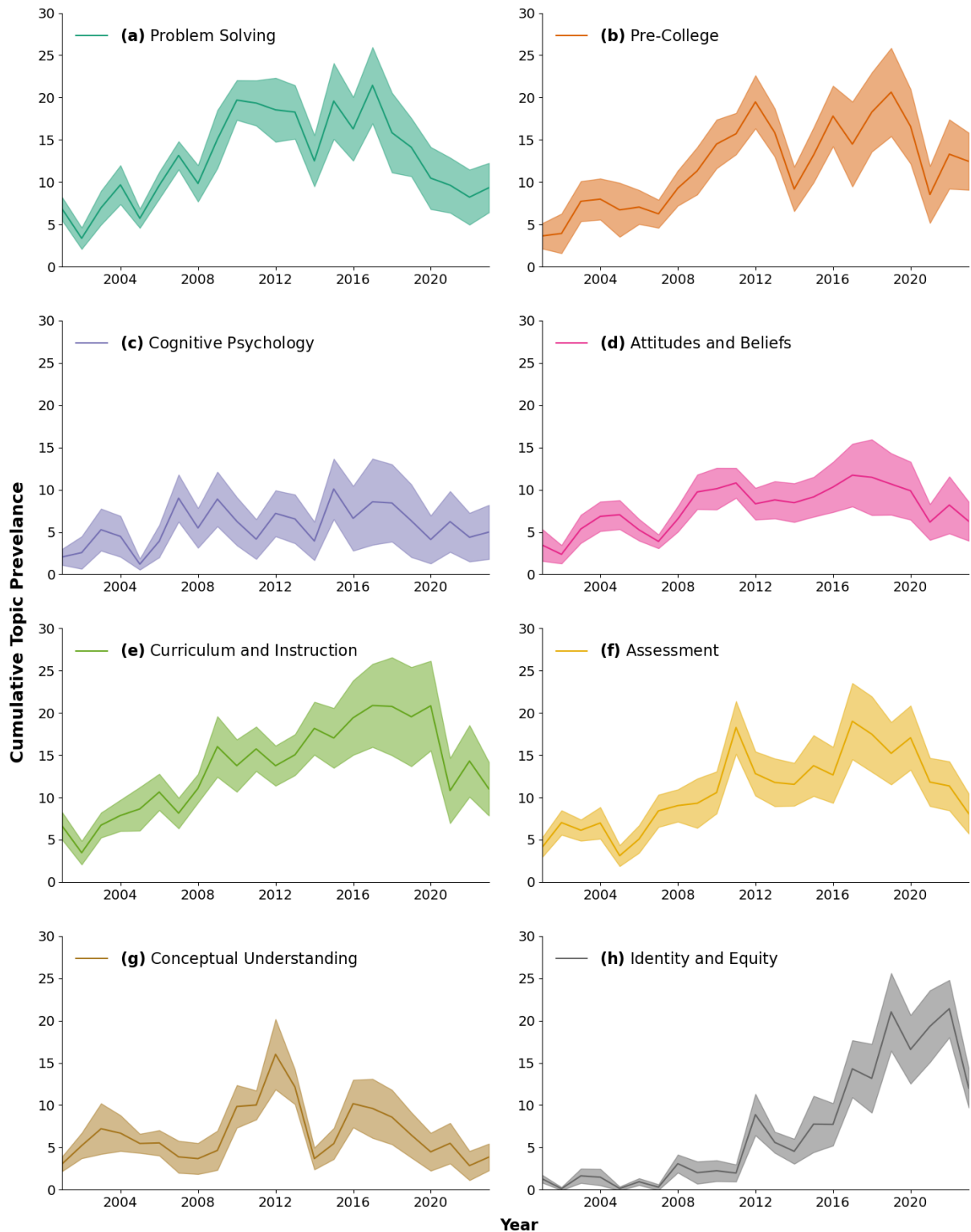

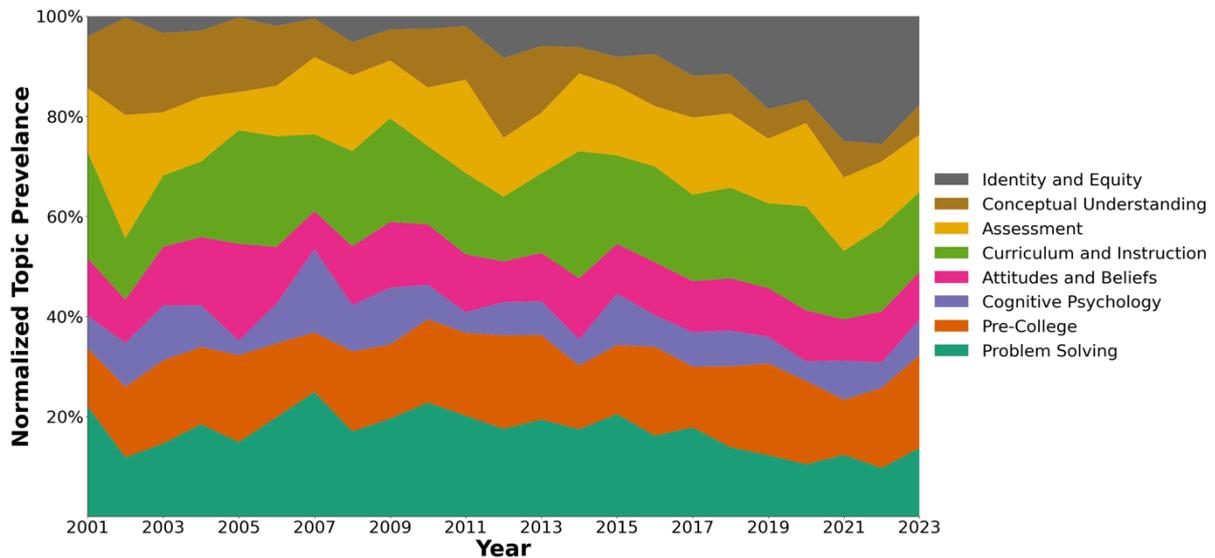

**FIG. 12.** Cumulative topic prevalence over time for researcher-defined topics. Y-axis is equivalent to approximate number of complete articles published on each topic per year. Error bars are three standard deviations of topic prevalence spread based on centroid resampling.

**FIG. 13.** Embeddings-based calculation of normalized prevalence as a function of time in PERC proceedings literature using researcher-defined topics.

To put these shifts in context, Figure 13 shows an alternate view of topic prevalence over time, in which all topics have been normalized to 100% each year. This view controls for changes in conference attendance and publication, and allows one to see that all topics in the model have received some amount of consistent interest over time, albeit at different levels. Some (like Cognitive Psychology and Attitudes and Beliefs) have had low prevalence for most of the PERC publication history, suggesting that they might either be more niche topics or are more commonly used in combination with others, like eye-tracking as a method for studying problem-solving or student beliefs as an outcome of educational reforms. However, even when controlling for variations in article publication numbers, certain topics have seen major surges or increases: Assessment appears to have had a small surge in the early years of the PERC proceedings, and Identity and Equity has seen consistent yearly growth since around 2016, topping out at around 20% of the literature in 2022. In contrast, research on Conceptual Understanding appears to have waned significantly in recent years, although it is still represented in the literature.

What caused these changes? Some, like the growth in research on Identity and Equity and the decline of research on Conceptual Understanding, seem to represent long-term trends in the field. Such changes are expected—all research fields shift over time, including education research [25]. It is important to note, however, that these literature shifts have not occurred in a vacuum. The PERC proceedings is the product of a living community of researchers, whose work is dependent on support from universities and government foundations. The physics education research community has changed radically over the years of the PERC proceedings publication, going from a small field in the process of establishing itself in the early years to a recognized field with its own specific journals and conferences in the present day. Additionally, factors like editorial decisions, publication norms, and conference locations have also certainly affected publication numbers and topics. An interesting area of future research would be to check the shifts visible in figures 12 and 13 against historical records of funding initiatives and conference participant recollections.

## IV. Discussion

### A. Reflections on the method

Our first two research questions were as follows: *Can embeddings be used to deductively replicate prior inductive qualitative research on scientific literature like the PERC proceedings?* And, *can embeddings be used to perform alternative analyses of this literature, using fully researcher-defined categories?* We argue that the results of this analysis demonstrate that embeddings can, indeed, be deductively used to both replicate prior inductive analysis and perform a new analysis using researcher-defined categories. Using only 15 representative articles from each of the LDA-defined topics, we were able to recover most of the recognizable aggregate trends in the LDA model as well as approximately 70% of the specific topic scores. Furthermore, the article categorizations by both the LDA-based embeddings model and the embeddings model based on Docktor and Mestre's literature review both have substantial face validity.

Given these results, embeddings analyses seem to offer significant advantages over other NLP approaches for qualitative data analysis. They require significantly less data cleaning than BoW approaches like LDA, and do not have the same issues with stochasticity as those methods [24,25]. They also only require a small amount of a-priori categorization, in the form of choosing a few (10-15) representative articles for each topic centroid. In the present analysis, exploratory modeling suggested that more texts provide greater resolution, i.e., decreased JS divergence between LDA and embeddings; but, there also seemed to be a point of diminishing returns, where doubling the number of articles only provides small decreases in JS divergence. This suggests that researchers using this method may be able to reasonably define categories with only a few example texts, either taken from the dataset or constructed by the researcher.

Furthermore, the method behind this analysis was relatively straightforward: 1) embed texts; 2) define centroids based on a selected sample of texts that exemplify desired features; 3) calculate topic scores based on transformed distances, using scaling parameters to determine "mixedness" of the model; and 4) evaluate results of the model analysis based on face validity of topic scores and general trends. Each of these components can easily be modified or replaced: for instance, the large language model used to create the embeddings can easily be swapped out or changed (and many such models are available, for free, on open-source repositories like HuggingFace); texts used to define centroids can easily be added or removed; multiple distance measures can be used; transformation functions can be replaced, tuned, or modified; and data can easily be added to or removed from the model without any negative downstream effects. Additionally, embeddings do not require nearly the same magnitude of data cleaning as bag of words-based techniques like LDA; as long as texts have been cleanly digitized, they can often be embedded and analyzed as-is.

Given this flexibility, we suggest that this technique holds great promise for helping researchers perform deductive qualitative analysis at scale on a wide variety of different types of data. Beyond literature reviews of research articles, one could easily imagine applying this same approach to code student responses on open-ended surveys or responses to conceptual physics questions, especially those that have already been partially analyzed by a researcher to define robust categories. One could also imagine segmenting interview transcripts and applying this type of analysis to different turns-of-talk or conversational segments in order to detect student ideas, attitudes, or misconceptions, similar to Sherin's (2013) analysis. In this regard, this embeddings-based approach could act as a tool for both automating existing analyses and drawing new inferences from data. One could even imagine multiple cycles of analysis in which edge cases like those described above are automatically detected (for example, by looking for topic "smear") and inspected by a human researcher, in order to modify or refine a coding scheme. This technique thus offers the opportunity to more flexibly integrate the capabilities of machine learning methods and human researchers, in line with calls by researchers like Kubsch et al. (2023) [43].

### B. What does this analysis tell us about the field of physics education research?

Returning to our third research question, *what do these analyses tell us about the development of the literature over time?* we can use these three analyses make some arguments about the development of the PERC proceedings literature, and arguably the state of American PER, as a function of time. First, all three analyses of the PER literature suggest that research on identity, equity, and learning communities has become an increased focus of the field. A survey of recent PERC conference themes supports this conclusion: for instance, "Working together to Strengthen the PER Community of Practice" (2023), "Queering Physics Education" (2022), "Making Physics More Inclusive and Eliminating Exclusionary Practices in Physics" (2021), "Insights, Reflections, & Future Directions: Emergent Themes in the Evolving PER Community" (2020), and "Physics Outside of the Classroom: Teaching, Learning, and Cultural Engagement in Informal Physics Environments" (2019). However, the present analysis allows us to make some quantitative estimates on the approximate amount of attention this topic is getting from the field: approximately 20-30%, depending on the model and whether communities of practice are combined with the topics of equity and identity. These trends appear to mirror trends in other related fields, like science education research, which has also seen a major shift towards research on equity, identity, and communities of practice in recent decades [25].

Second, all three analyses show that all topics have seen some amount of sustained interest over time, lasting up to the present day. Although some topics (like Cognitive Psychology) seem to be more niche, and some (like Conceptual Understanding) have waned over time while others (like Identity and Equity) have grown, it seems clear that all of the topics described above remain interesting to the research community, and new researchers entering the field have a wide variety of intellectual traditions and empirical results on which to build. At the same time, given the fact that theoretical perspectives on equity, identity, and communities of learning have typically been imported into PER from other fields like sociology, psychology, and the learning sciences, their rise suggests that researchers interested in defining new research paradigms would do well to familiarize themselves with theories and traditions outside the field. This kind of intellectual "cross-pollination" can be a rich source of new ideas and approaches for studying physics teaching and learning [25].

### C. Limitations and Future work

We have shown a proof-of-concept that embeddings can be used for deductive qualitative research. As discussed, other researchers have shown that embeddings can also be used for inductive qualitative research, such as computational grounded theory [31].

However, a great deal of work remains to understand how best to leverage embeddings in qualitative data analysis. For example, this study was done with an out-of-the-box, general-purpose large language model; such models can be domain-adapted or fine-tuned to specific contexts, which allows them to better capture (and distinguish) the meaning of texts from those contexts. One could easily imagine domain-adapting a model using literature from studies of physics teaching and learning (such as articles from *The Physics Teacher*, *Physics Education*, or *The American Journal of Physics*) to provide better resolution on PER-specific language and ideas.

Further, there is a need to critically evaluate the way embeddings capture the meaning of texts. For instance, is a single vector enough to capture meaning from a text, or might one gain better insight by representing different parts of a text using multiple vectors? What are the benefits and tradeoffs of different vector dimensionalities for different types and lengths of text? We also see a need to explore the space of different distance measures and transformations. The transformations used in this analysis were fairly straightforward, primarily meant to invert and magnify minute differences in distance in a high-dimensional meaning space. There are certainly other more sophisticated transformations that may give better results for specific applications, especially when combined with different distance measures. In this regard, we can count ourselves lucky to be part of a field that has significant experience with using vectors to analyze the behavior of physical systems.

In the future, order to critically evaluate this method these types of analyses will need to be benchmarked against existing qualitative analyses, using standard statistical methods for evaluating inter-rater reliability. One downside of the present analysis is the fact that the embeddings analysis was benchmarked against another NLP model, and neither model was benchmarked against any kind of "ground truth" (in the form of human-performed qualitative analysis). This, we feel, was an appropriate approach for a proof-of-concept (and offers interesting insights into the differences between LDA and embeddings for topic analysis) but it leaves open the question of how such models would perform relative to a human coder. This suggests that a natural next step would be to apply this technique to large, text-based datasets that have already been qualitatively coded, like those from Wilson et al. (2022) [19] or Sabo et al. (2016) [6], and compare the results to human codes.

Finally, because this method is fundamentally a form of qualitative analysis it inherits many of the limitations inherent to those family of methods. That is to say, the analytic choices the researcher makes (for example, which categories to include or which example texts should be used to define topic centroids) will have a significant effect on the results. Furthermore, the results of this type of analysis require some amount of further qualitative interpretation, like inspecting topic scores of articles to determine face-validity and adjusting model parameters when results do not make sense. There are certainly benefits to this NLP approach to qualitative analysis, in that it is scalable to large datasets and reproducible. However, there are new potential sources of bias and issues with transparency: for example, although the general method for turning texts into vectors is well understood at a technical level, the actual details of the process will vary based on the particular large-language model used and researchers will need to carefully evaluate if the models they are using are capturing the particular semantic meanings they care about in the texts. It may be that addressing these issues will require some kind of combination of traditionally qualitative and quantitative analysis methods.

## V. CONCLUSION

Qualitative research has long been a time-consuming task, wherein the researcher themselves essentially acts as the instrument for analysis [44]. Computational tools have been useful for managing datasets and handling the mechanics of the process, but their limited ability to extract meaning from text greatly constrained what they were able to do. Now, we as a field have reached a point where we can use AI to quantitatively analyze and compare meaning in text. This technological development will continue; already, companies are connecting large language models to image models, so soon it may even be possible to analyze meaning across multiple representational modes. Thus, the future of qualitative research will likely look radically different from what we, as a field, have known so far. It is time that we, as a field, begin to explore the use of these new tools. We look forward to seeing what is possible, and hope that these tools will soon help us to gain a deeper and broader understanding of physics teaching and learning.

## ACKNOWLEDGMENTS

We gratefully acknowledge the contribution of Alessandro Marin in developing the methods that led to this work. We would also like to thank Danny Caballero, Ben Zwickl, David Hammer, Noah Finkelstein, Valerie Otero, Kirsty Dunnett, Emily Bolger, and Cassandra Lem for their thoughts and feedback on this work. This work was funded by the Norwegian Agency for International Cooperation and Quality Enhancement in Higher Education (NOKUT) which supports the University of Oslo's Center for Computing in Science Education and the Center for Interdisciplinary Education.

## DATA AVAILABILITY STATEMENT

The code and data used in this analysis are openly available on GitHub, at https://github.com/markusuio/G.E.V.I.R, and Zenodo at https://zenodo.org/records/10702781